\documentclass[%
aip,
amsmath,amssymb,
reprint,
]{revtex4-1}

\usepackage{graphicx}
\usepackage[percent]{overpic}
\usepackage[]{pdfpages}

\graphicspath{ {./fig/} }

\usepackage{xcolor}

\makeatletter
\AtBeginDocument{\let\LS@rot\@undefined}
\makeatother

\begin{document}

\title{Uncertainty Quantification of First Principles Computational Phase Diagram Predictions of Li-Si System Via Bayesian Sampling }

\author{Ying Yuan}
\affiliation{Department of Materials Science and Engineering, Carnegie Mellon University, Pittsburgh, Pennsylvania 15213, USA}
\author{Gregory Houchins}
\affiliation{Department of Physics, Carnegie Mellon University, Pittsburgh, Pennsylvania 15213, USA}

\author{Pin-Wen Guan}
\affiliation{Department of Mechanical Engineering, Carnegie Mellon University, Pittsburgh, Pennsylvania 15213, USA}

\author{Venkatasubramanian Viswanathan}
\affiliation{Department of Materials Science and Engineering, Carnegie Mellon University, Pittsburgh, Pennsylvania 15213, USA}
\affiliation{Department of Physics, Carnegie Mellon University, Pittsburgh, Pennsylvania 15213, USA}
\affiliation{Department of Mechanical Engineering, Carnegie Mellon University, Pittsburgh, Pennsylvania 15213, USA}

\begin{abstract}

Within the field of computational materials discovery, the calculation of phase diagrams plays a key role. Uncertainty quantification for these phase diagram predictions enables a quantitative metric of confidence for guiding design in computational materials engineering. In this work, an assessment of the CALPHAD method trained on only density functional theory (DFT) data is performed for the Li-Si binary system as a case study. with applications to the modeling of Si as an anode for Li-ion batteries. Using a parameter sampling approach based on the Bayesian Error Estimation Functional (BEEF-vdW) exchange-correlation. By using built-in ensemble of functionals from BEEF-vdW, the uncertainties of the Gibbs Free Energy fitting parameters are obtained and can be propagated to the resulting phase diagram. To find the best fitting form of the CALPHAD model, we implement a model selection step using the Bayesian Information Criterion (BIC) applied to a specific phase and specific temperature range. Applying the best selected CALPHAD model from the DFT calculation, to other sampled BEEF functionals, an ensemble of CALPHAD models is generated leading to an ensemble of phase diagram predictions. The resulting phase diagrams are then compiled into a single-phase diagram representing the most probable phase predicted as well as a quantitative metric of confidence for the prediction. This treatment of uncertainty resulting from DFT provides a rigorous way to ensure the correlated errors of DFT is accounted for in the estimation of uncertainty.  For the uncertainty analysis of the single-phase diagram of the Li-Si system, we explore three different methods using BEEF as three kinds of samplers with various assumptions of statistical independence: independent points of phases, independent pairs of phases, and independent convex hulls of phases. We find that each method of propagating the uncertainty can lead to different phases being identified as stable on the phase diagram. For example, the phase Li$_{4.11}$Si at 300K is predicted to be stable by all functionals using the second and third method, but only 15\% of functionals predict it to be stable using the first method. From the phase diagram, we have also determined intercalation voltages for lithiated silicon. We see that the same phase can have a distribution of predicted voltages depending on the functional. In combination, we can generate a better understanding of the phase transitions and voltage profile to make a more analysis-informed prediction for experiments and the performance of Si-anodes within batteries.

\end{abstract}

\maketitle

\section*{Introduction}
The prediction of phase diagrams and phase transformations are important for many energy applications, especially in Li-ion batteries.\cite{persson2010thermodynamic}  Computational prediction of equilibrium phase diagrams using density functional theory has been used successfully to predict the thermodynamics of intercalation electrodes.\cite{persson2010thermodynamic,pande2018robust}  However, the prediction of equilibrium phase diagrams involves various sources of uncertainty: uncertainty associated with numerical predictions of density functional theory, choice of the exchange correlation functional,\cite{wellendorff2012density}  uncertainty associated with the choice of a model for describing the thermodynamics of the system such as cluster expansion \cite{ruban2008configurational}  and uncertainty associated with fitting the parameters of the chosen model.\cite{choi2008inductive}


Quantifying the uncertainty is important as this could lead to vastly different conclusions on the identified stable phases and the associated thermodynamics.\cite{houchins2020towards}  The challenge associated with systematic uncertainty quantification and propagation through a model has limited the application of these methods to calculation of phase diagrams (CALPHAD).  There have been approaches proposed for uncertainty quantification within CALPHAD.\cite{stan2003bayesian,otis2017high}  Stan et al., proposed a weighted genetic algorithm sampling tool to estimate the posterior probability of a free energy model parameters.  Otis and Liu performed model selection of a CALPHAD model using both Akaike Information Criterion (AIC) and F-test.  Using this in conjunction with a Monte-Carlo sampling scheme determined the posterior probability distribution.  In a recent work, Honarmandi et al. performed a thorough evaluation of uncertainty of CALPHAD model parameters and the resulting phase diagrams of Hf-Si binary system.\cite{honarmandi2019bayesian}  A large source of uncertainty within these predictions stems from the choice of the exchange correlation functional for DFT calculated data.\cite{wellendorff2012density,decolvenaere2015testing,kitchaev2016energetics,lejaeghere2014error}  Bayesian error estimation capabilities of the BEEF-vdW exchange correlation function has been used to quantify uncertainty associated with a variety of DFT-predicted material properties.\cite{PhysRevB.96.134426,PhysRevB.94.064105,Parks2019Uncertainty,guan2019uncertainty}

In this work, we implement model-parameter selection through the use of Bayesian Information Criterion (BIC) applied to specific phases and specific temperature range to find the optimal number of parameters and parameter types in the CALPHAD models. 
\begin{table}
\centering
\label{tab:1}
\caption{Phases in the Li-Si binary system: Phase names and crystallographic data accepted in this work and experimental data reported in literature.}
\medskip
\begin{tabular}{ccc}
\hline
Phase\;name & Pearson\;symbol/Space  & Reference \\ & group/Lattice\;parameter(pm) & \\
\hline
Li$_{17}$Si$_4$ & cF420 & [\citenum{zeilinger2013single}] \\
 & F$\bar{4}$3m & \\
  & a = 1872.59(1) & \\
Li$_{22}$Si${5}$ & cF432 & [\citenum{axel1966kenntnis}] \\
 & F23 & \\
  & a = 1875.0 & \\
Li$_{21}$Si$_{5}$ & cF416 & [\citenum{nesper1987li21si5}] \\
 & F$\bar{4}$3m & \\
 & a = 1871.0 & \\
Li$_{4.11}$Si & orthorhombic & [\citenum{zeilinger2013revision}] \\
 & Cmcm & \\
 & a = 452.46(2) & \\
 & b = 2194.4(1) & \\
 & c = 1320.01(6) & \\
Li$_{15}$Si$_4$ & cI76 & [\citenum{zeilinger2013stabilizing}] \\
 & I$\bar{4}$3d & \\
 & a = 1063.22(9) & \\
Li$_{13}$Si$_4$ & oP34 & [\citenum{zeilinger2013}] \\
 & Pbam & \\
 & a = 794.88(4) & \\
 & b = 1512.48(8) & \\
 & c = 446.61(2) & \\
Li$_7$Si$_2$ & oP34 & [\citenum{schafer1965kristallstruktur}] \\
 & Pbam & \\
 & a = 799 & \\
 & b = 1521 & \\
 & c = 443 & \\
Li$_7$Si$_3$ & hR7 & [\citenum{von1980struktur}] \\
 & R$\bar{3}$m & \\
 & a = 443.5(1) & \\
 & c = 1813.4(3) & \\
Li$_2$Si & mS12 & [\citenum{axel1965kristallstruktur}] \\
 & C2/m-$C^3_{2h}$ & \\
 & a = 770 & \\
 & b = 441 & \\
 & c = 656 & \\
Li$_{12}$Si$_7$ & oP152 & [\citenum{nesper1986li12si7}] \\
 & Pnma & \\
 & a = 860.0 & \\
 & b = 1975.5 & \\
 & c = 1433.6 & \\
LiSi & tI32 & [\citenum{tang2013synthesis}] \\
 & I 4$_1$/a($n^o88$) & \\
 & a = 935.45(5) & \\
 & c = 573.74(5) & \\
\hline
\end{tabular}
\end{table}
An implementation of Bayesian Error Estimation with van der Waals correction, the BEEF-vdW \cite{wellendorff2012density} was trained on a set of benchmark data spanning solid-state properties,  covalent organic systems, noncovalent and van der Waals interactions, and chemisorption on solid metallic surfaces. A functional form for the exchange-correlation potential is fit through the least-squares fitting of the error in predictions for this training set. The parameters space can then be perturbed slightly to sample a collection of models that are marginally above the minimum of the least-squares fit and therefore should provide a nearly equally good fit. 
Within this work, we use the ensemble of functionals from BEEF as a sampler for the generation of CALPHAD models in order to propagate the uncertainty from DFT to the final prediction of phase diagrams. We assume that every phase diagram of different functionals from BEEF can provide potentially useful information as it samples the predictions of a point in exchange-correlation space. The BIC model selection method is applied for each chosen solid phase in our Li-Si binary system and a specific temperature range of 200K-450K.  We implement a Bayesian Information Criterion (BIC) model selection step to determine best model from a 10-parameter CALPHAD model space.   Using this approach, we not only explore the method of using only DFT to make a phase prediction but also give the quantitative range of uncertainty using DFT data at the GGA-level.

\section*{Li-Si Binary System}

Li-ion batteries have played an important role in electrifying transporation.\cite{goodenough2013li, thackeray2012electrical}  Next-generation batteries are required for electrifying trucking\cite{sripad2017performance,Sripad2017EvaluationOpportunities} and aviation.\cite{fredericks2018performance,Viswanathan2019}  Silicon anodes are a promising candidate among next-generation chemistries to improve the energy density of batteries.\cite{mcdowell201325th}  Hence, a detailed and consistent thermodynamic description of the Li-Si binary system is critical for silicon anode development.

 
 Uncertainty quantification of first principles computational phase diagram predictions via Bayesian sampling can help achieve reasonable phase predictions of Li-Si system and a full understanding of phase transitions during charging and discharging. In this work, we used the set of phases in the Li-Si binary system as in Table  I \ref{tab:1}, which compiles the crystallographic data information of the solid phases including Li$_{17}$Si$_4$\cite{zeilinger2013single}, Li$_{22}$Si$_5$\cite{axel1966kenntnis}, Li$_{21}$Si$_{5}$\cite{nesper1987li21si5}, Li$_{4.11}$Si\cite{zeilinger2013revision}, Li$_{15}$Si$_4$\cite{zeilinger2013stabilizing}, Li$_{13}$Si$_4$\cite{zeilinger2013}, Li$_7$Si$_2$\cite{schafer1965kristallstruktur}, Li$_7$Si$_3$\cite{von1980struktur}, Li$_2$Si\cite{axel1965kristallstruktur}, Li$_{12}$Si$_7$\cite{nesper1986li12si7}, LiSi\cite{tang2013synthesis}, as well as solid-phase BCC Li, solid-phase Diamond Si as previously used in work by Lang \textit{et. al.} 2017.\cite{Liang2017thermodynamics}

\begin{figure}
    \centering
    \includegraphics[width = 0.48\textwidth]{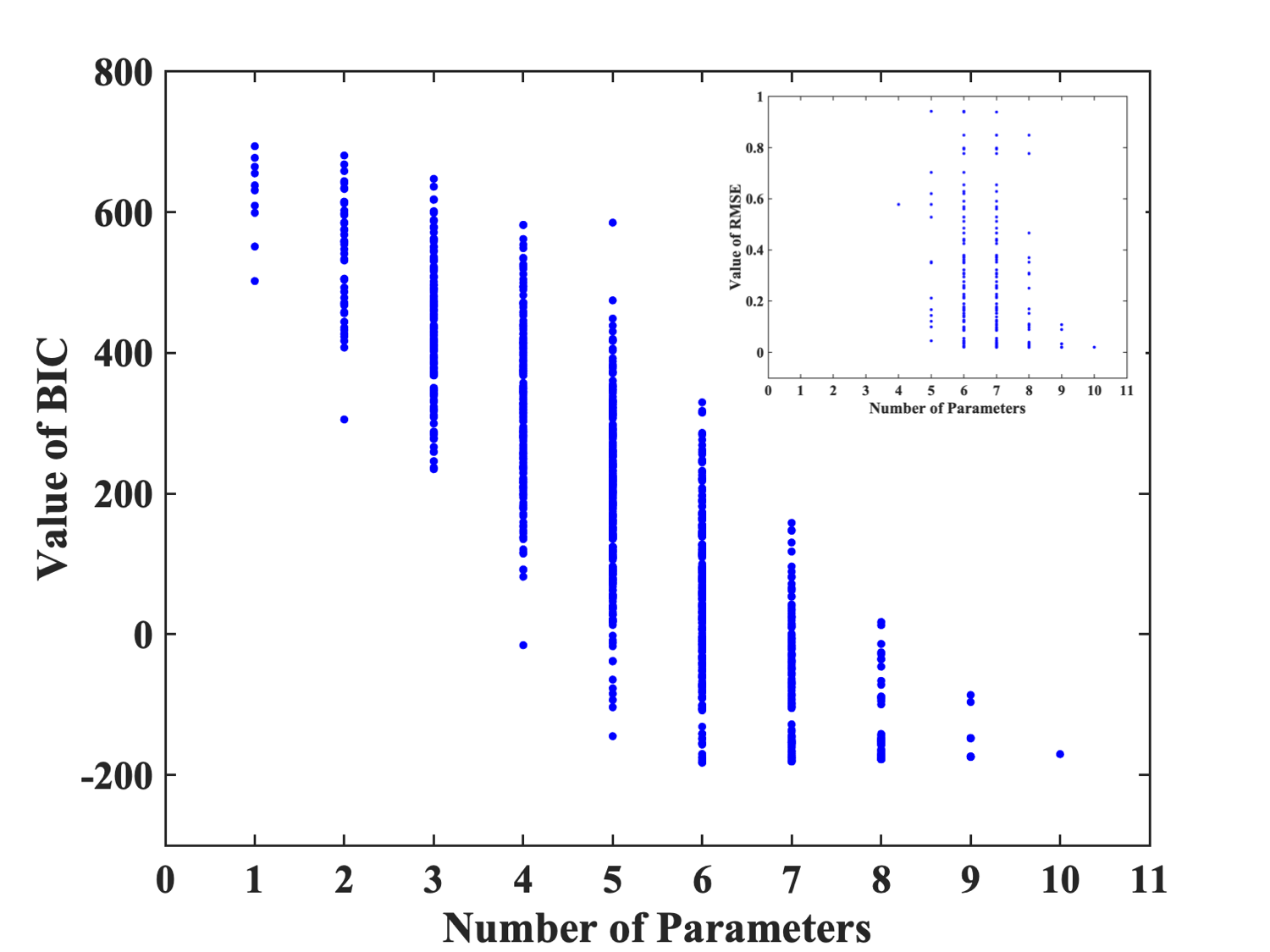}
    \caption{BIC model selection results for solid phase Li$_{15}$Si$_4$. The inset picture shows the results of RMSE corresponding to different models.}
    \label{fig:1}
\end{figure}
 
\section*{Methods}
To compute the energy of each structure, the Projector Augmented Wave method of Density functional theory, as implemented in GPAW \cite{enkovaara2010electronic}, was used. The exchange correlation potential was treated at the genearalized gradient approximation level using he Bayesian Error Estimation Functional with van der Waals (BEEF-vdW)\cite{wellendorff2012density}. For each structure, the atomic positions were first relaxed to a maximum force of 0.03 eV/\AA, and the computed energy at various volumes was fit to a vinet equation of state\cite{vinet1987compressibility} using either 7 or 9 different volumes depending on the difficulty of fitting the equation of state for the ensemble of functionals.

\begin{table*}
\centering
\label{tab:2}
\caption{The BIC model selection results of Li-Si system in temperature range 200K-450K. The complexity of model structure of Gibbs energy function is 10.}
\medskip
\begin{tabular}{ccc}
\hline
phase\;name  \quad & \quad BIC\;model\;selection\;result \quad & \quad number\;of\;parameters \\
\hline
Li & a+bT+d$T^2$+e$T^{-1}$+g$T^7$+i$T^4$+jlnT & 7 \\
Si & a+bT+cTlnT+e$T^{-1}$+jlnT & 5 \\
Li$_{15}$Si$_4$ & a+bT+cTlnT+e$T^{-1}$+i$T^4$+jlnT & 6 \\
Li$_{22}$Si$_{5}$ & a+bT+cTlnT+d$T^2$+f$T^3$+jlnT & 6 \\
Li$_{21}$Si$_5$ & a+bT+cTlnT+d$T^2$+f$T^3$+jlnT & 6 \\
Li$_{13}$Si$_4$ & a+bT+cTlnT+e$T^{-1}$+g$T^7$+jlnT & 6\\
LiSi & a+bT+cTlnT+d$T^2$+f$T^3$+g$T^7$+i$T^4$ & 7 \\
Li$_7$Si$_2$ & bT+cTlnT+d$T^2$+e$T^{-1}$+f$T^3$+jlnT & 6 \\
Li$_2$Si & a+bT+cTlnT+d$T^2$+jlnT & 5 \\
Li$_{12}$Si$_7$ & bT+cTlnT+d$T^2$+d$T^{-1}$+f$T^3$+i$T^4$+jlnT & 7 \\
Li$_{17}$Si$_4$ & a+bT+cTlnT+d$T^2$+f$T^3$+jlnT & 6 \\
Li$_{4.11}$Si & a+cTlnT+d$T^2$+f$T^3$+g$T^7$+i$T^4$ & 6 \\
Li$_7$Si$_3$ & a+bT+cTlnT+d$T^2$+i$T^4$+jlnT & 6 \\

\hline
\end{tabular}
\end{table*}


Using the fitted properties of the equation of state, a Debye-Grunessen theory analysis\cite{moruzzi1988calculated} was used to incorporate vibrational properties and predict the Gibbs free energy as a function of temperature as the Debye model is a reasonable approximation that yields finite temperature thermodynamics of sufficient accuracy\cite{guan2019uncertainty}.  This process was repeated for the ensemble of 2000 non-self consistent exchange-correlation functionals within the BEEF-vdW model space framework, but not every functional can lead to successful result (see computational details of Supplimentary Information). 

Next, to best fit the predicted Gibbs energy curve for each phase, we employ model selection through the use of BIC. The BIC was chosen as it tends to penalize complex models more heavily, giving preference to simpler models in selection than the Akaike information criterion (AIC)\cite{hastie2005elements} especially for a small number of training points. The Gibbs energy of the pure element $i$ in phase $\Phi$, namely the Gibbs energy function used within CALPHAD modeling, is adopted in this work as the following equation \cite{wang2013thermodynamic}:
\begin{multline}
G^{0,\Phi}_i(T)=G^\Phi_i(T)-H^{SER}_i=a+bT+cTlnT+dT^2\\
+eT^{-1}+fT^3+gT^7+hT^{-9}+iT^4+jlnT
\end{multline}
In this equation, the Gibbs energies of the pure solid (Li)-bcc, (Si)-diamond and all stoichiometric compounds are described with absolute reference state. $H^{SER}_i$ is the molar enthalpy of the element (i) at 298.15 K and 1 bar in its standard element reference (SER) state. T is the absolute temperature. There are 10 possible parameters, corresponding to different functional dependencies on T. While the physical meaning of each parameter is largely empirical, here, parameter a relates to the static energy, parameter b describes the entropy, and parameters c, d, e, f, g, h, i and j are determined from the temperature dependence of the heat capacity, $C_p$, for each compound: 
\begin{multline}
C_p=-c-2dT-6eT^2-2fT^{-2}-42gT^6-90hT^{-10}\\-12iT^3+jT^{-1}
\end{multline}
Although each parameter has its relation to a specific physical meaning, for fitting part of the Gibbs energy result, it is not necessary to choose all of these parameters to get the best fitting equation\cite{lukas2007computational}. For the full 10-parameter model space, we iterate through all unique combinations of these 10 parameters for a total of $2^{10}-1=1023$ different G-T fitting models tested for each phase. 
BIC is defined as \cite{Schwarz1978estimating} 
\begin{align}
BIC(M)&=k\log(n)-2\log(\bar{L})\\
&=k\log(n)+n*\log(\frac{\sum(Y_i-f_i(X))^2}{n})
\end{align}
where k is the number of parameters of each model, $\bar{L}$ is the maximized likelihood function, n is the number of training points of each model, here we take n as 26 for each model and each phase in temperature range 200K-450K. $Y_i$ is the true value of one training point, $f_i$(X) is the value of prediction after G-T fitting by using the model. For the results of BIC values of different models for each phase, choosing the model with minimum BIC is equivalent to choosing the model with the largest (approximate) posterior probability\cite{hastie2005elements}, the model with the lowest BIC value is predicted to be the most ideal model, namely the model that best trades off accuracy and model complexity.

\begin{figure}
    \centering
    \includegraphics[width=0.48\textwidth]{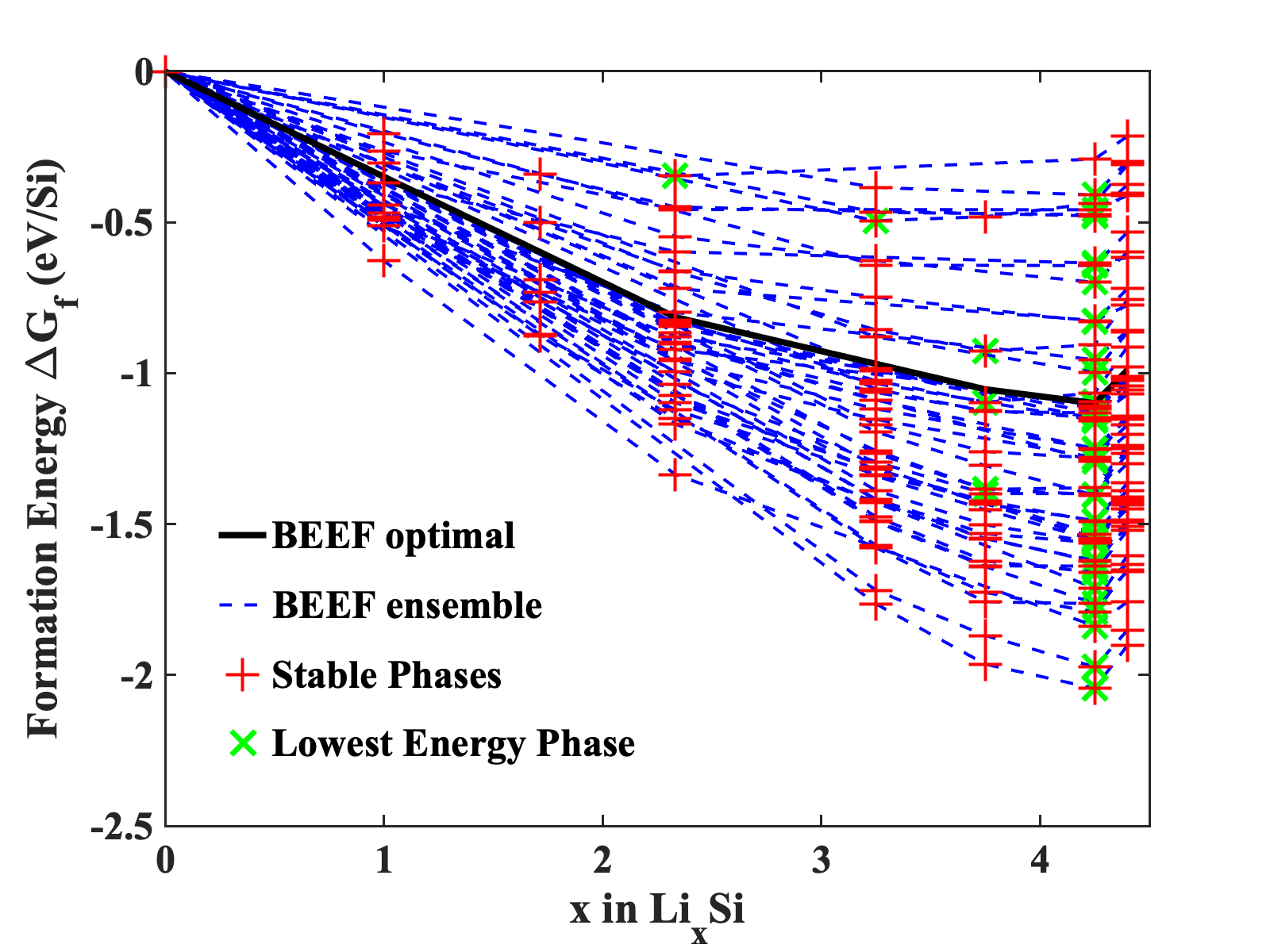}
    \caption{A collection of 50 predictions from the full ensemble of convex hulls is shown is shown in blue.  The BEEF optimal function is also shown as a black solid line. The stable phases appear in each convex hull are labeled with red cross. The lowest energy phase of each convex hull is labeled with green x, which represents the maximum stable lithiated silicon phase predicted by each functional.}
    \label{fig:2}
\end{figure}

After determining the best fitting model for each phase by applying BIC criterion to the data from the optimal BEEF functional, we fit the Gibbs energy curve for each phase and the ensemble of functionals. The results are then stored in a thermodynamic database (TDB) file as is conventional in CALPHAD. These TDB files are then read into the pycalphad software \cite{otis2017pycalphad} to get phase diagram data of each functional and replotted to produce the uncertainty phase diagram for the Li-Si system automatically, which is a new method to achieve automating first-principles phase diagram calculations different from A. van de Walle and G. Ceder's former work\cite{van2002automating}, another automated algorithm.

In CALPHAD, the uncertainty can be propagated from the model parameters to Gibbs free energy, then to the phase diagram. In previous work incorporating uncertainty within phase diagram predictions, as discussed above, Markov Chain Monte Carlo (MCMC) sampling approach has conventionally been used to obtain plausible optimum values and uncertainties of the parameters \cite{honarmandi2019uncertainty,honarmandi2019bayesian}. This propagation of uncertainty assumes that the errors in the prediction of two materials from DFT are independent of one another. It is well known, however, that the errors of DFT predictions contain correlated, systematic prediction errors. Thus, within this work, we utilize BEEF as a sampler of GGA-level exchange-correlation potential space to understand the correlated uncertainty of the underlying ab initio thermodynamics data and its effect on the final prediction. Different functionals of BEEF-vdW lead to different predictions of both enthalpy from DFT and vibrational properties from the Debye analysis, and thus different value of G-T fitting parameters. This will then result in various predictions of stable phases. By propagating this uncertainty from the different functionals to the model parameters, then to the Gibbs free energy and ultimately the phase diagram, we can then assign a quantitative prediction confidence to the result.

In this work, we also considered the computation of equilibrium cell voltages based on the thermodynamics\cite{urban2016computational,pande2018robust}. The electrochemical lithium-coupled ion transfer reaction with silicon is given by:
\begin{multline}
x(Li^++e^-)+Si \rightleftharpoons Li_xSi
\end{multline}
The Gibbs free energy change associated with this reaction is then given by:
\begin{multline}
\Delta{G} = G_{Li_xSi}-G_{Si}-xG_{Li^+}-xG_{e^-}
\end{multline}
where G$_{Li_xSi}$ is the free energy of the given Li-Si phase, G$_{Si}$  is the free energy of the solid pure silicon phase, G$_{Li^+}$ is the free energy of the Li-ion including the energy of solvation with the electrolyte and G$_{e^-}$ is the free energy of the electron at the potential of Si electrode. For ease, 
we can relate the sum of the free energy of the Li-ion and the electron to the free energy of bulk lithium given by the reaction:
\begin{multline}
Li^++e^- \rightleftharpoons Li_{(s)}
\end{multline}
and therefore $G_{Li^+} + G_{e^-_{U=0V}} = G_{Li_{(s)}}$. This sets the zero potential to the $Li/Li^+$ redox potential within the given electrolyte so $G_{e^-} = G_{e^-_{U=0V}} -eU_{Li/Li^+}$. We finally can write the change in Gibbs energy as:
\begin{multline}
\Delta{G} = G_{Li_xSi}-G_{Si}-xG_{Li_{(s)}}+x(eU_{Li/Li^+})
\end{multline}
and compute the the intercalation potential of a particular Li-Si phase. Additionally we can derive the  intercalation potential by considering the phase transformation from a phase  Li$_{x_1}$Si to another phase Li$_{x_2}$Si given by:
\begin{multline}
U_{Li/Li^+} = \dfrac{-1}{e(x_2-x_1)}(G_{Li_{x_2}Si}-G_{Li_{x_1}Si}-(x_2-x_1)G_{Li_{(s)}})
\end{multline}
Thus the important quantity needed is the free energies of the stable phases and from this, the itercalation potential can also be determined. This free energy not only contains the enthalpy which can be estimated as the internal energy given from DFT, but also contains vibrational contributions in the was of entropy and zero point energy. $\Delta{G} = \Delta{H}-T\Delta{S}+\Delta{ZPE}$. 

To estimate the vibrational properties of the Gibbs energy relating to the zero-point energy and entropy, a Debye-Grunessen theory analysis \cite{moruzzi1988calculated} was performed using the DePye software \cite{guan2019uncertainty} which enables the efficient processing and vibrational predictions of the ensemble of functionals. 

\begin{figure}[htbp]
		\begin{minipage}[t]{0.9\linewidth}
			\begin{overpic}[width=1.0\linewidth,trim={0 -0.5cm 0 0cm},clip]{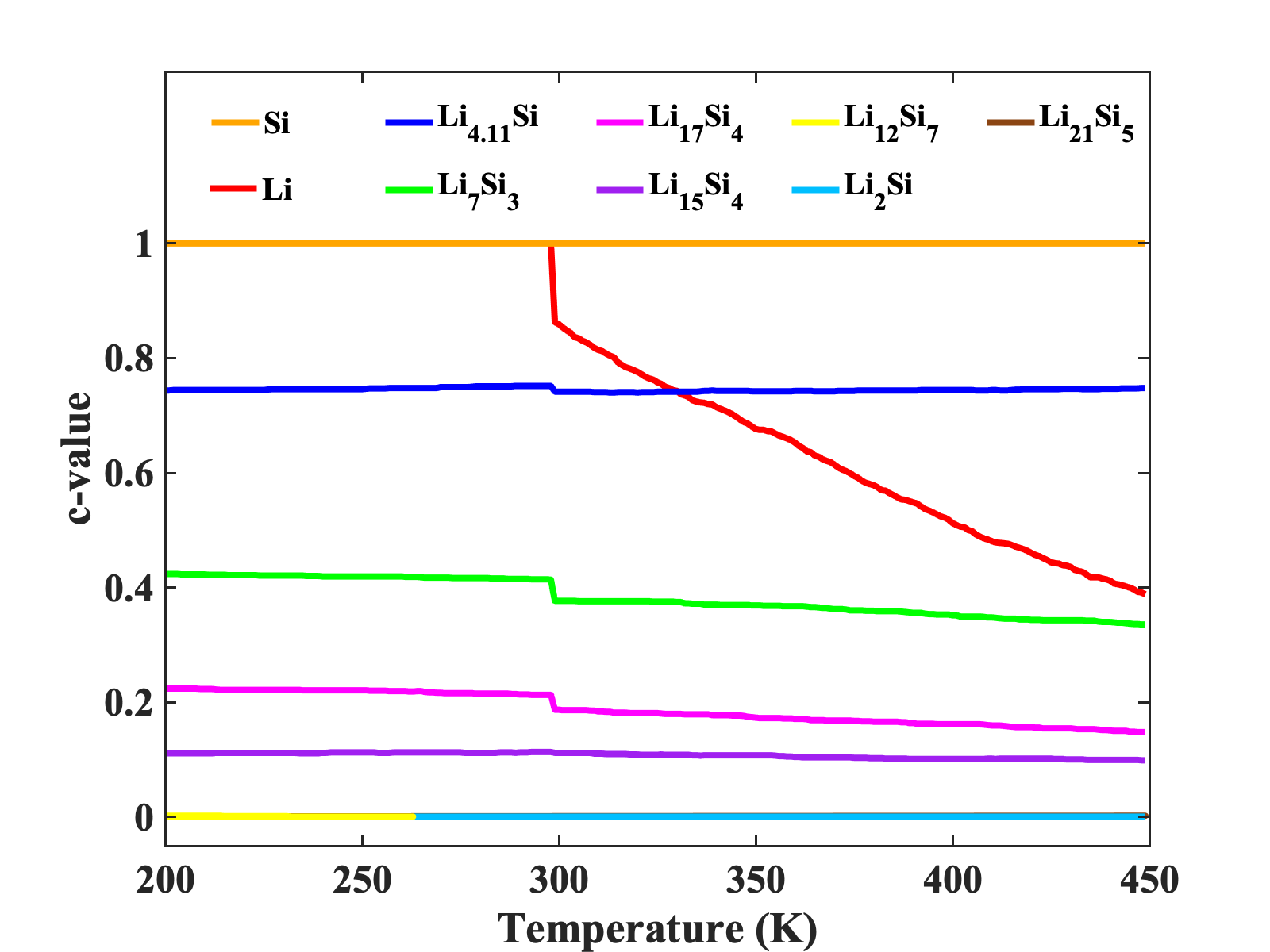}
				\put (-0.15,5) {\colorbox{white}{\large{\textbf{(a)}}}}
			\end{overpic}
		    \begin{overpic}[width= 1.0\linewidth,trim={0 -0.5cm 0 0cm},clip]{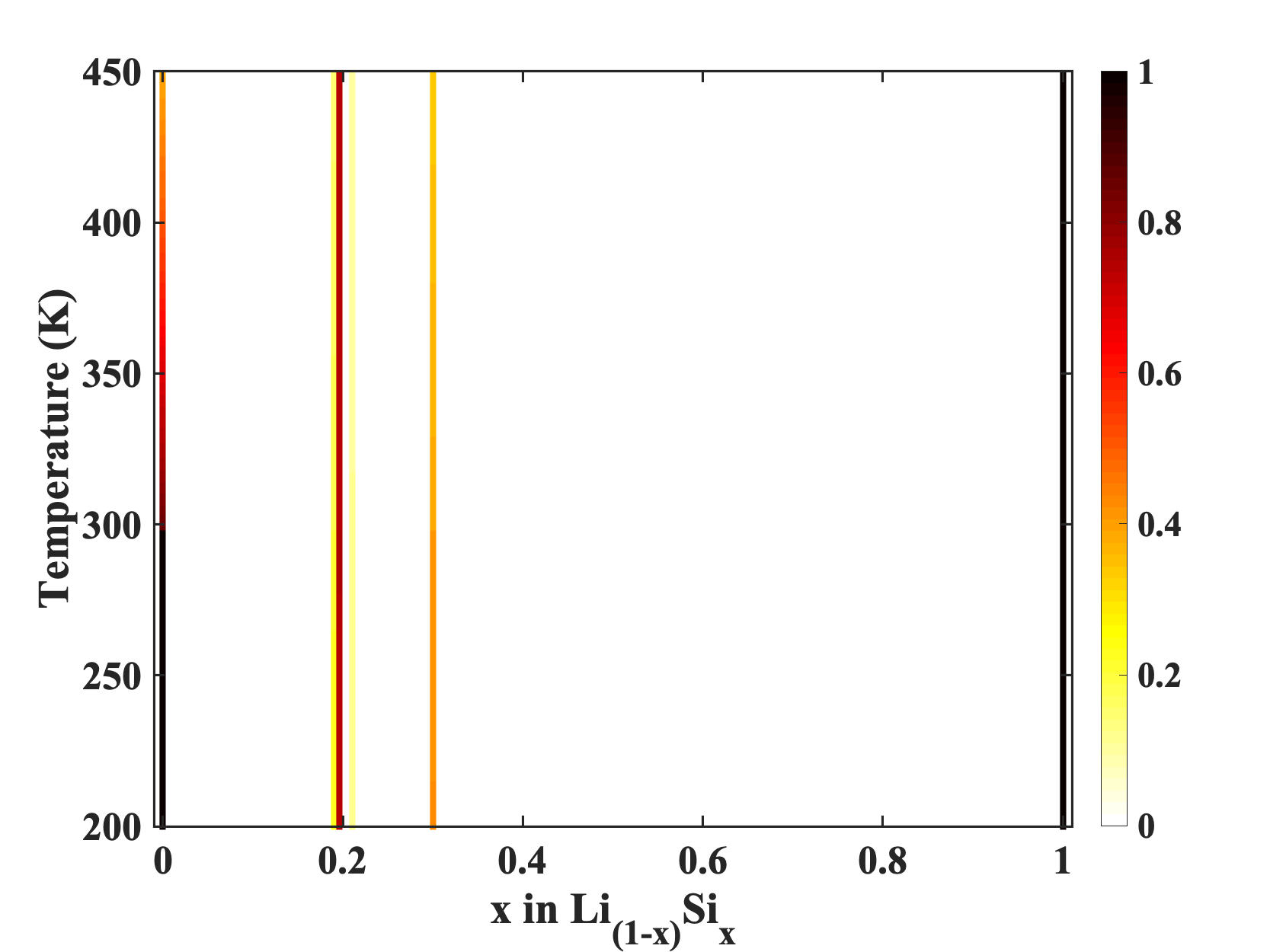}
				\put (-0.15,5) {\colorbox{white}{\large{\textbf{(b)}}}}
			\end{overpic}
		    \begin{overpic}[width= 1.0\linewidth,trim={0 -0.5cm 0 0cm},clip]{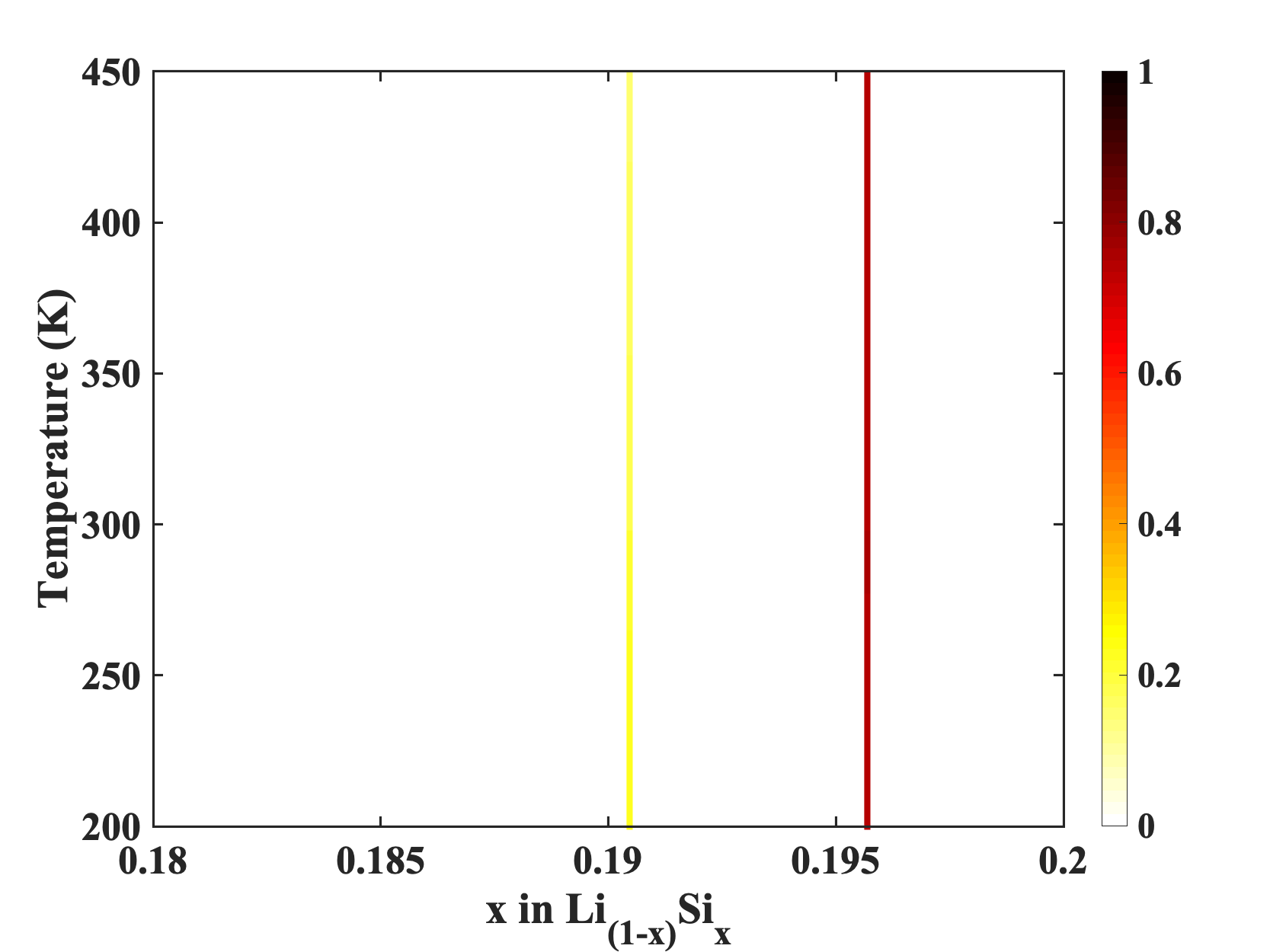}
				\put (-0.15,5) {\colorbox{white}{\large{\textbf{(c)}}}}
			\end{overpic}
		\end{minipage}		
		
		\caption{\label{fig:3} Single phase stability in Li-Si system: (a) c-value of different singles phases as a function of temperature. (b) Contour map of confidence-value for single phases as a function of temperature and (c) enlarged version for the region, $0.18 < x < 0.2$ in Li$_{(1-x)}$Si$_x$.}
\end{figure}

\section*{Results and Discussion}


A sample BIC model selection for Li$_{15}$Si$_4$ within the 10-parameter model space for the temperature range 200K-450K is shown in Figure \ref{fig:1}, with the BIC model selection of all other materials available in the Supplementary Information. In this case, the lowest value of the BIC corresponds to a 6 parameters model, while for other materials the number of parameters varied from 5 to 7 as can be seen in Table II\ref{tab:2} along with the specific model chosen.  Although the number of parameters chosen is consistent with the number commonly used in the literature about CALPHAD, different phases have different specific combinations of parameters rather than a fixed set of parameters as is conventionally used. The results show that although a higher number of parameters is not necessarily useful in practice, the specific terms used are important for the goodness of fit and thus will provide a more accurate and reasonable phase diagram prediction. The inset picture shows the results of root mean squared error (RMSE) of the prediction of training data corresponding to different models of solid-phase  Li$_{15}$Si$_4$. For this plot and all corresponding plots for RMSE in the SI, we see that the error continues to decrease as the number of parameters increases while there is a distinct minimum for the BIC as it is expected that as the number of model parameters increases, over-fitting will occur.  Thus BIC allows for the selection of a model with both a small error and a small number of parameters.

\begin{figure}
		\begin{minipage}[t]{0.95\linewidth}
			\begin{overpic}[width=1.0\linewidth,trim={0 -0.5cm 0 0cm},clip]{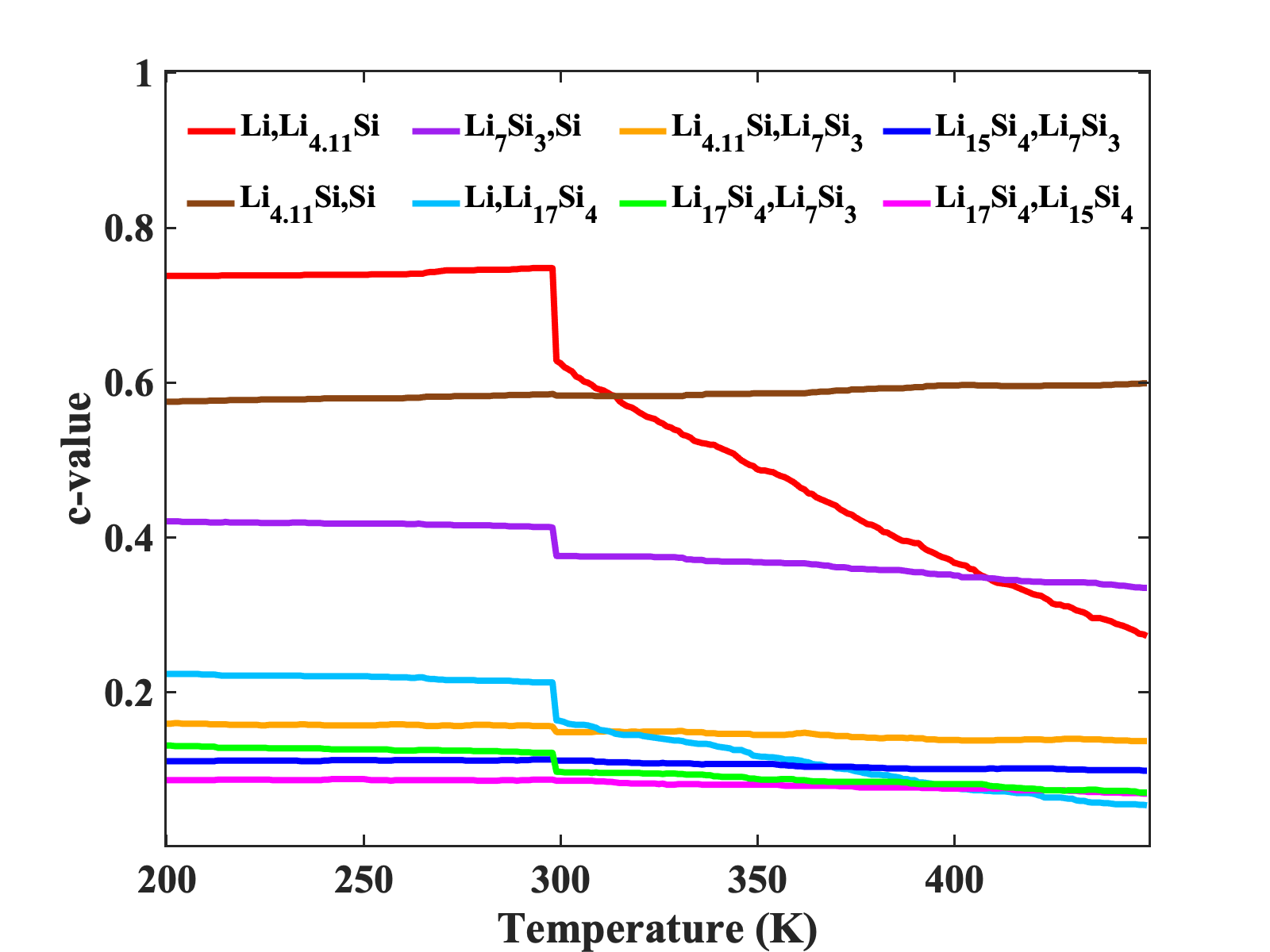}
				\put (-0.15,5) {\colorbox{white}{\large{\textbf{(a)}}}}
			\end{overpic}
		    \begin{overpic}[width= 1.0\linewidth,trim={0 -0.5cm 0 0cm},clip]{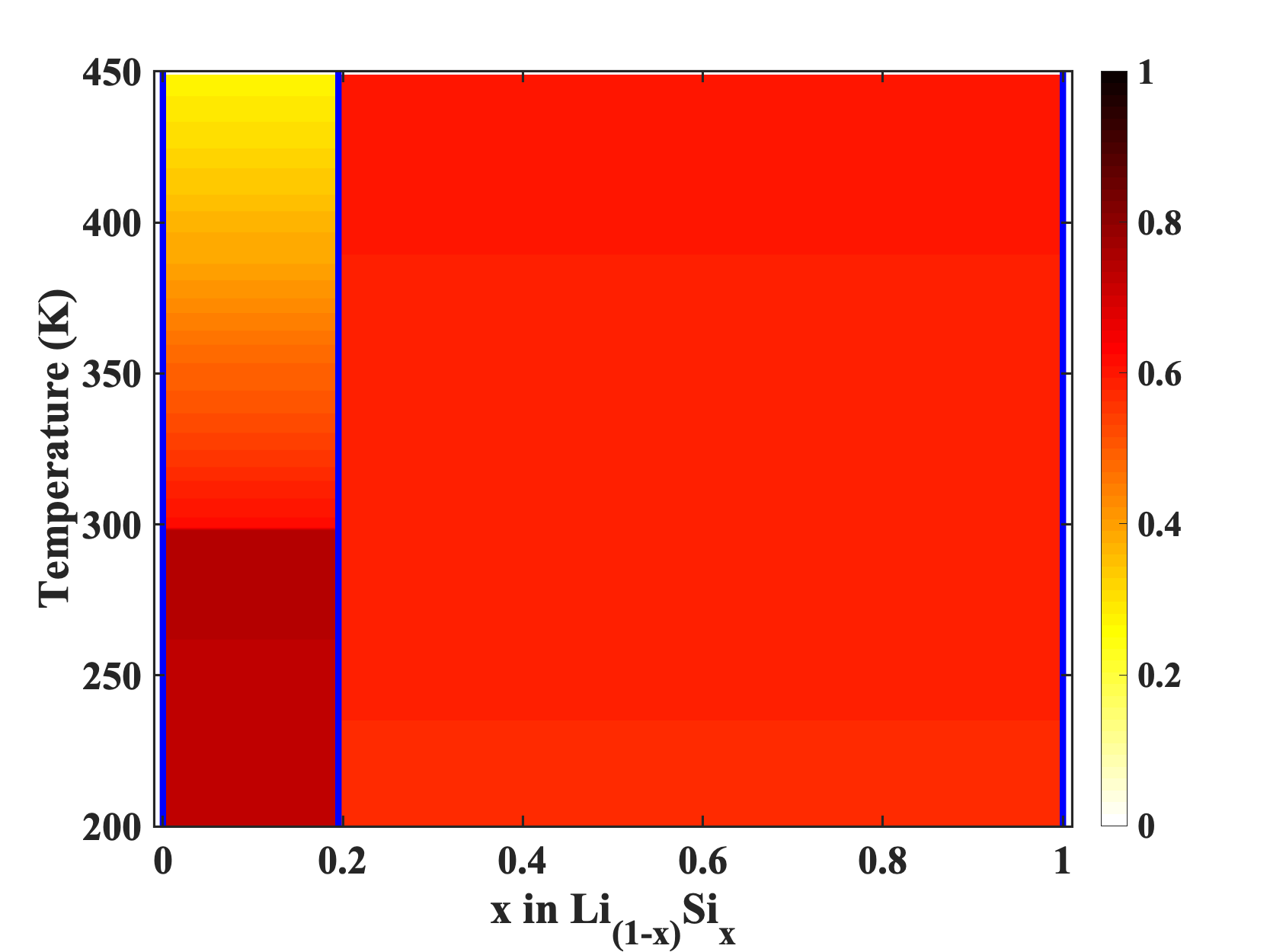}
				\put (-0.15,5) {\colorbox{white}{\large{\textbf{(b)}}}}
			\end{overpic}
		\end{minipage}		
		
		\caption{\label{fig:4} Stability of pairs of phases in Li-Si system: (a) c-value of pairs of phases as a function of temperature. (b) Contour map of confidence-value for pairs of phases as a function of temperature.}
\end{figure}

\begin{figure}
		\begin{minipage}[t]{0.95\linewidth}
			\begin{overpic}[width=1.0\linewidth,trim={0 -0.5cm 0 0cm},clip]{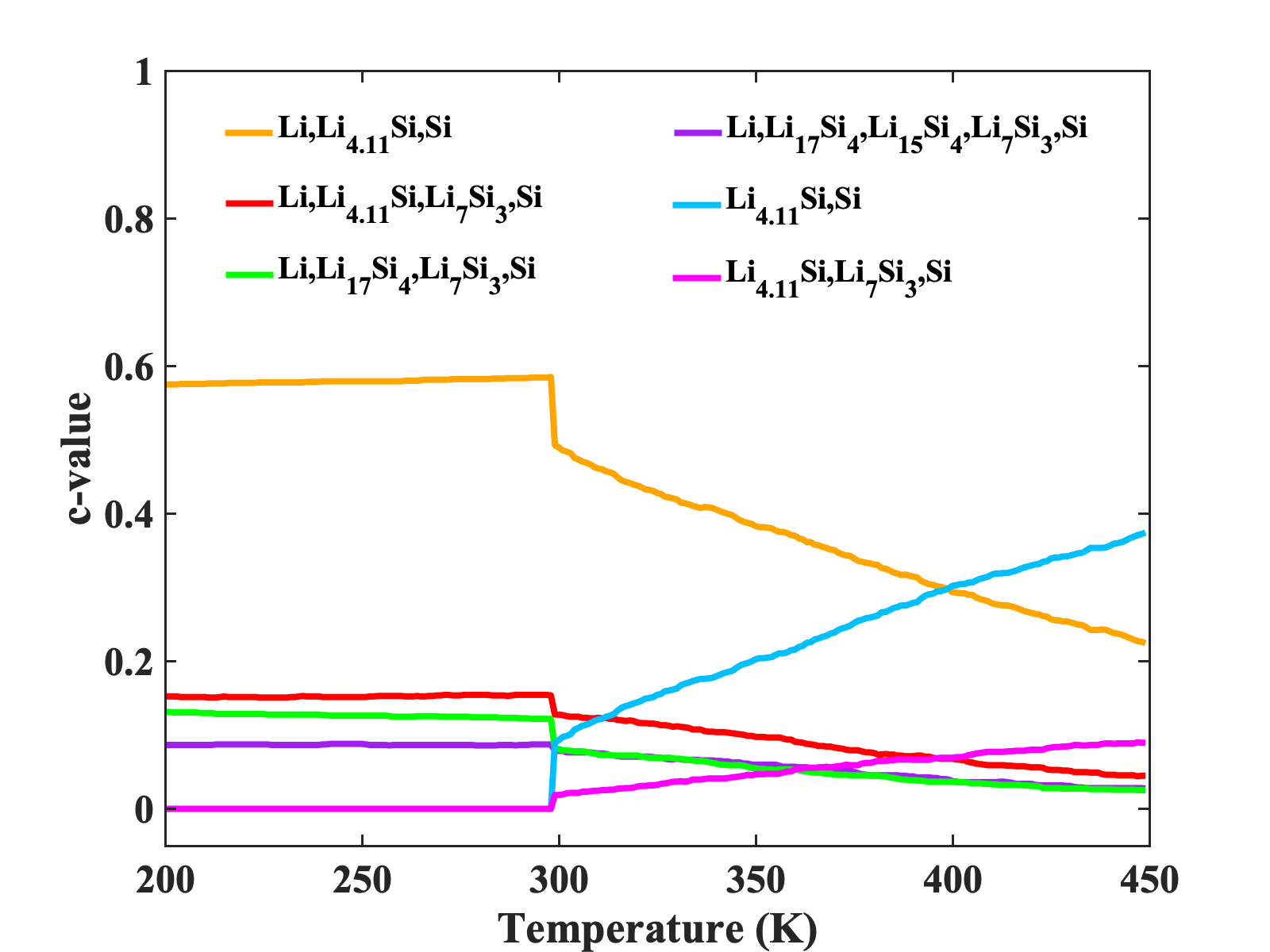}
				\put (-0.15,5) {\colorbox{white}{\large{\textbf{(a)}}}}
			\end{overpic}
		    \begin{overpic}[width= 1.0\linewidth,trim={0 -0.5cm 0 0cm},clip]{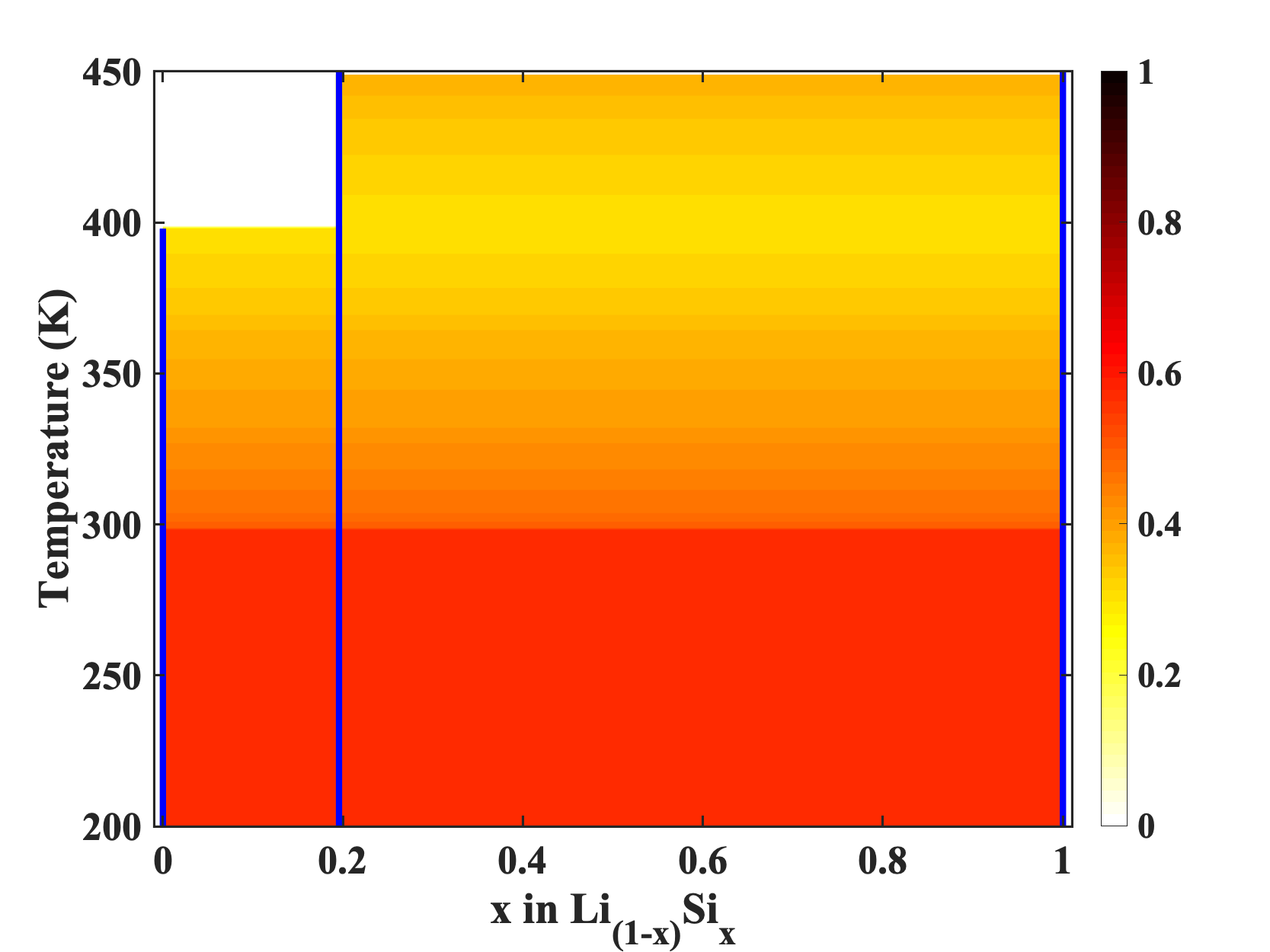}
				\put (-0.15,5) {\colorbox{white}{\large{\textbf{(b)}}}}
			\end{overpic}
		\end{minipage}		
		
		\caption{\label{fig:5} Convex hulls of phases for the Li-Si system: (a) c-value of specific convex hulls as a function of temperature.  (b) Contour map of confidence-value for specific convex hulls as a function of temperature.}
\end{figure}





The phase diagram is determined by the structures on the convex hull of the free energy vs x in Li$_x$Si diagram.  We show a sampling of the convex hull identified by 50 exchange correlation functionals within the BEEF-vdW model space at 300 K in Figure \ref{fig:2}.  We find that many of the GGA functionals predict similar convex hulls, while some predict vastly different stable phases and formation energies.  Further, even for functionals having the same convex hull, the energy differences are vastly different, which will influence the intercalation voltages. The features are similar to that identified in an earlier work of the Li-C phase diagram by Pande et al.\cite{pande2018robust}  This shows the importance of the exchange correlation functional in determining the phase diagram.\cite{lenchuk2019comparative}


As mentioned previously, we apply the results of this BIC model selection from the 10-parameter model space from optimal BEEF functional into other functionals of the ensemble for the temperature range is between 200K-450K. For the uncertainty analysis in phase diagram, we explore three different methods using BEEF as three kinds of samplers with various assumptions of statistical independence: independent points of phases, independent pairs of phases, and independent convex hulls of phases.

For the first case in which we assume the probability of prediction of a phase is independent of the prediction of any other phase, we define the c-value as the normalized number of times a functional predicts a phase to occur at a given temperature. That is the number of times a particular phase appears in the ensemble of phase diagrams. The specific c-value of each phase for each temperature is shown in Figure \ref{fig:3}(a). Figure \ref{fig:3}(b) and (c), shows this c-value plotted as a function of composition and temperature where the larger the c-value, the darker the color is. It can be seen that at a fixed temperature of 300K in Figure \ref{fig:3}(b), not all phases appear with high c-value of near 1. Rather some phases are predicted to be stable with higher confidence than others. Here we regard the confidence value beyond 0.05 as a good prediction value. In Figure \ref{fig:3}(a), phases Li$_{12}$Si$_7$, Li$_2$Si, and Li$_{21}$Si$_5$ are predicted to be unstable with confidence close to 0. In comparison with the phase diagrams from literature built on experimental data\cite{Liang2017thermodynamics} where phases only appear or do not appear, this incorporation of uncertainty allows for a more reliable prediction given the uncertainties present and the sensitivity of those uncertainties on the final result. There are many influencing factors and errors in experimental data. However, in the uncertainty phase diagram, we consider all of the different prediction results from the BEEF ensemble so that we can predict the occurrence probability of one phase point. At one phase with increasing temperature, if c-value also increases, the phase is stable at high temperatures, otherwise, the phase is stable at low temperatures. For example, we can say that solid Li is stable at low temperatures and Li$_{12}$Si$_7$ is very unlikely to be stable according to the c-value of the UQ phase diagram. In the previous uncertainty phase diagram found in literature,\cite{honarmandi2019uncertainty,honarmandi2019bayesian} only the uncertainty range of a specific phase is given. Therefore, in this uncertainty phase diagram, it can not only give the uncertainty range but also give the occurrence probability compared with the uncertainty phase diagrams in P. Honarmandi et al's work \cite{honarmandi2019bayesian}. Hence, this kind of uncertainty phase diagram can help determine a better phase prediction and understand the process of phase transition of a system better. 

When we consider the composition of mixing phases or the correlation between two adjacent predicted stable phases even all predicted stable phases, the former method is not enough, especially when trying to predict the exact phase mixture at compositions that are not a distinct phase. To do this, we utilize  two additional methods that consider the occurrence of pairs together, as well as the occurrence of a the full set of phases on a hull. For the pairs of phases case in Li-Si system, similarly, we define the c-value as the normalized number of times a functional predicts a pair of phases to occur at a given temperature. Though there are 13 phases in Li-Si system, there are just 22 kinds of valid pairs collected form each convex hull, after removing the pairs with small probability(less than 0.05), there are just 8 kinds of valid pairs to be considered. Among them, there are just 2 kinds of pairs show with max probability when x in range 0 to 1 and T in range 200 to 450K. The specific c-value of each pair of phases for each temperature is shown in Figure \ref{fig:4}(a). Figure \ref{fig:4}(b) shows the corresponding phase diagram. Blue lines represents the phase boundaries of pairs of phases. For the phases between the phase boundaries, the deeper the color, the bigger the c-value. We can see when temperature is below 300K, phase with x in range between 0 and 0.196 will be predicted as the mixing of solid Li phase and solid Li$_{4.11}$Si phase with the maximum probability 0.75. With temperature increasing, this prediction will not change but the maximum c-value will decrease, whose physical meaning is that the solid Li phase may disappear because of the melting point. As for a phase with x in the range between 0.196 and 1.0, it is predicted to be the mixing of solid Li$_{4.11}$Si phase and solid Si phase with the maximum c-value 0.6 or a bit higher during the temperature range from 200 to 450K.

For the case of comparing final predicted convex hulls in Li-Si system, we define the c-value as the normalized number of times a functional predicts a specific convex hull of phases to occur at a given temperature. That is the number of appearance times of a convex hull of phases in the ensemble of phase diagrams divided by the number of functionals of an ensemble. Though there are 13 phases in Li-Si system, there are just 27 kinds of valid convex hulls predicted by the functionals of the ensemble, after removing convex hulls with a small probability(less than 0.05), there are just 6 kinds of the valid convex hull to be considered. Among them, there are just 2 convex hulls shown with max probability when x in range 0 to 1 and T in range 200 to 450K. The specific c-value of each convex hull of phases for each temperature is shown in Figure \ref{fig:5}(a). Figure \ref{fig:5}(b) shows the corresponding phase diagram. The crossing points of each horizontal T line and blue lines represent the predicted stable phases belong to the convex hull appearing with maximum probability at that temperature. For the phases between blue lines, the deeper the color, the bigger the c-value. We can see when the temperature is below 400K, the convex hull including stable solid phase Li, solid Li$_{4.11}$Si, and solid Si will be predicted appearing with the maximum c-value. Specifically, this prediction stays as 0.58 when T is below 300K, but with temperature increasing, the c-value will decrease, as the solid Li phase is predicted to disappear because of melting. When T increases from 400K to 450K, the convex hull including stable solid phase Li$_{4.11}$Si and solid Si will be predicted appearing with the maximum c-value but the value decreases. As for phase with x in range between 0.196 and 1.0, it is predicted to be the mixing of solid Li$_{4.11}$Si phase and solid Si phase with equal c-value of the convex hull it belongs to. The physical meaning of the white area is that the liquid phase Li may appear with the same c-value.

\begin{figure}
	\centering
	\includegraphics[width = 0.48\textwidth]{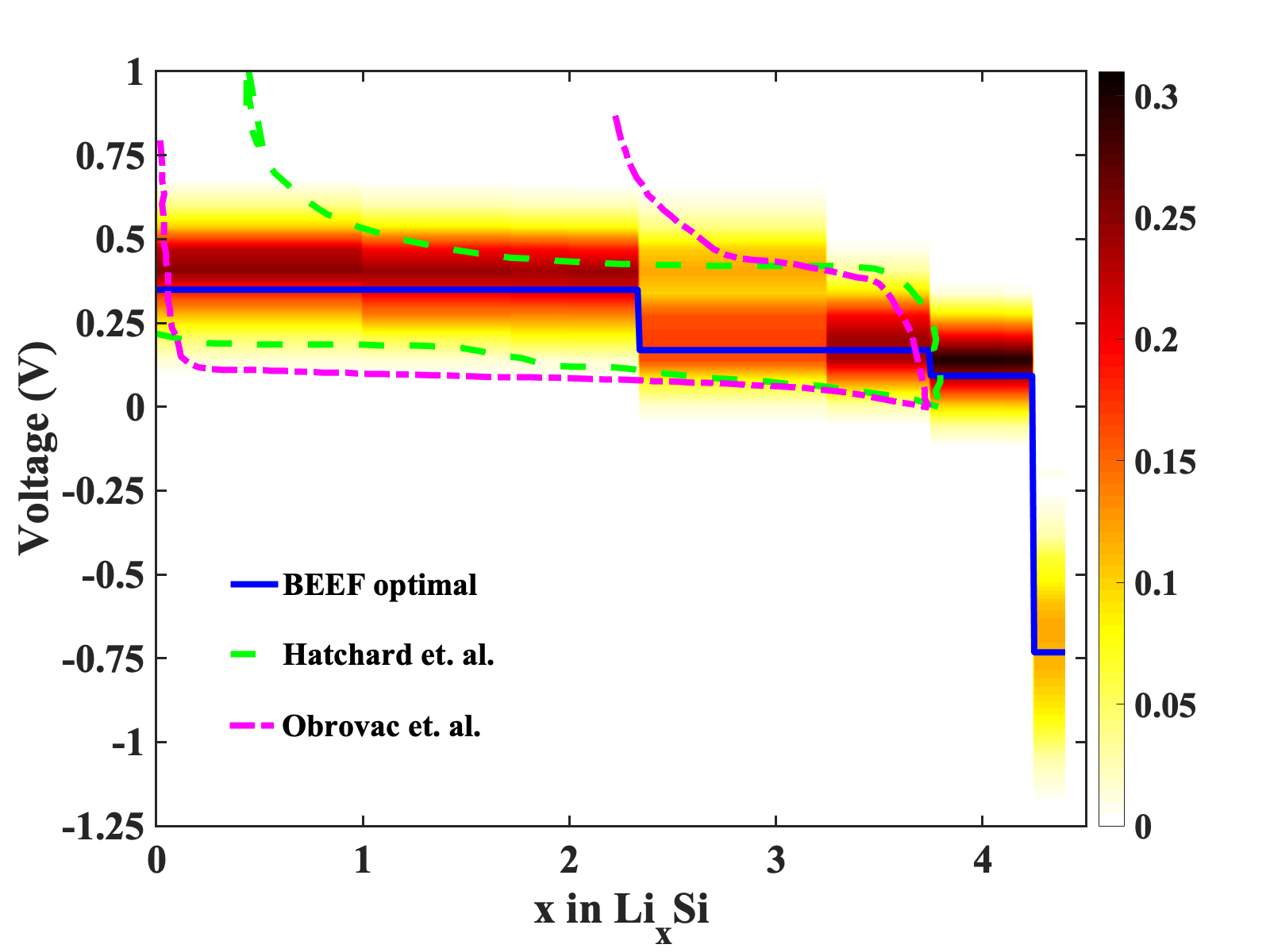}
	
	\caption{\label{fig:6} Voltage prediction for the Si-anode system. At each x in Li$_x$Si, we show the probability density function as a countour and the optimal BEEF prediction as a solidblue line represents. The green and pink lines are from the experimental measurements of T.D. Hatchard et al. and M.N. Obrovac et al.\cite{hatchard2004situ,obrovac2004structural}}
\end{figure}

From the previous convex hull, we then evaluate the intercalation potentials, using the predicted energy of stable phases. The ensemble of energy predictions then generates and ensemble of intercalation potentials. In order to visualize these various predictions, the results at each composition were binned into a histogram with 0.066 V bin width. This then generates a probability distribution function at every composition with is then plotted as the  contour map in Figure \ref{fig:6}. From this figure, we see the maximum intercalation potential predicted with maximum probability varies slightly from the the BEEF optimal but give the range that includes the experimental data well. We can see here, different functionals can not only predict different phase transition but also predict different energy difference, both of them will influence the prediction of voltage. The same phase transition (namely the same pair of phases) may have huge voltage prediction difference because of the huge difference of energy difference between two phases from different functionals. And different phase transition (namely different pairs of phases) may have very close voltage prediction. If we just consider the c-value of phase, we may get better phases transition prediction; if we just consider the probability density of voltage, we may get better voltage profile prediction. By using both in combination, we can have a better understanding of the phase transitions and voltage profile to make a better prediction for experiments and the performance of the Li-Si system within batteries.

\section*{Conclusion}

In this work, we employed model selection using the BIC to determine the best model from a list of independently generated models for a specific phase and specific temperature range using DFT-data and the Debye-Grunessen model.  A sampling of GGA-space within the DFT data is carried out using the built-in error estimation capabilities of the BEEF-vdW exchange correlation function.  Using this, we determine the uncertainty associated with Li-Si binary system, an important candidate Li-ion battery anode.  We carry out three different approaches to uncertainty quantification for the Li-Si phase diagram. These three methods include various levels of statistical correlation between the prediction of phases.  This analysis provides a basis to further extend uncertainty quantification of first principles data into the phase diagram predictions. We believe that quantifying the uncertainty will provide a more detailed assessment of the possible phase diagram and one particular use is to identify the regions of largest uncertainty to guide the most useful experiment to be done for the most information gain related to the phase diagram. 



\begin{acknowledgments}
		Y.Y would like to thank Dilip Krishnamurthy and Olga Vinogradova for their insightful input. G.H gratefully acknowledges funding support from the National Science Foundation under award CBET-1604898. Acknowledgment is also made to the Extreme Science and Engineering Discovery Environment (XSEDE) for providing computational resources through award number TG-CTS180061.
\end{acknowledgments}

\bibliography{cite}
\includepdf[pages={{},{},1,{},2,{},3,{},4,{},5,{},6,7}]{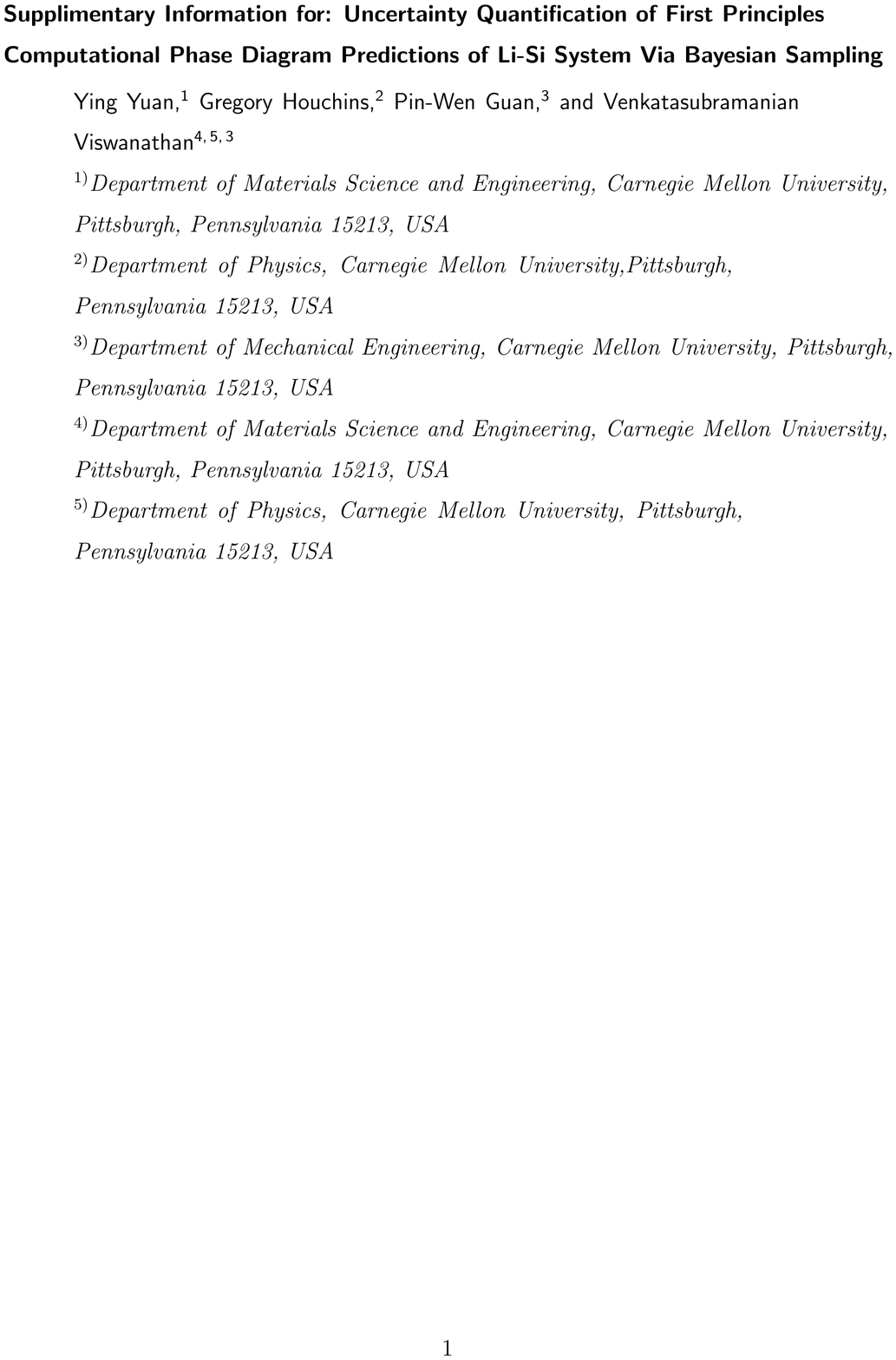}

\end{document}